\documentclass{aa}  
\usepackage{graphicx}
\usepackage{psfig}
\usepackage{txfonts}
\begin{document}
   \titlerunning{The soft X-ray spectrum  of Mrk 335}
   \authorrunning{A.L. Longinotti et al.}
   \title{The Seyfert 1 Galaxy Mrk 335 at a very low flux state: mapping the soft X-ray photoionised gas}


   \author{A.L. Longinotti, A. Nucita, M. Santos-Lleo, M.Guainazzi}

   \offprints{A.L. Longinotti}

   \institute{ESAC - European Space Astronomy Centre,  P.O. Box 78, 28691 Villanueva de la Ca{\~n}ada, Madrid, Spain     \\
              \email{alonginotti@sciops.esa.int}
                       }

   \date{Received January 2008;}

 
  \abstract
   {This paper reports on an XMM-Newton  observation of the Seyfert 1 Galaxy Mrk 335 performed as a target of opportunity when the source was in an unusually  low flux state.      
     }
   {The low level of continuum emission unveiled an underlying  line-rich soft X-ray spectrum, which can be  studied  with the  Reflection Grating Spectrometer. }
   {The emission features were analysed at high resolution with unprecedented detail for this source. Line ratio diagnostics from H-like and He-like ions indicate that  the line emission arises in  X-ray photoionised plasma.
    Extensive simulations were performed with the  {\small CLOUDY} photoionisation  code.
    The physical  properties of the line emitting material were derived from the comparison of the expected and observed line intensities.}
   { Because of  the  degeneracy in the ionisation parameter, a number of different solutions for the electron density and column density of the gas are consistent with the spectral diagnostics. 
   This prevents us from  uniquely  determining   the properties of the plasma; however, the location(s) of the X-ray photoionised gas can be constrained to within 0.06 pc (outer boundary). This limit places the X-ray line emitting gas at the inner edge of the material where the broad optical lines are produced (broad line region).
     }
   {}


   \maketitle
%

\section{Introduction}
The soft X-ray spectrum of Seyfert galaxies has represented  for astronomers a region of great interest and, at the same time, a sort of challenge. 
With the advent of high-resolution spectrometers onboard {\it XMM-Newton} and {\it Chandra},
it has been possible to reveal the nature of emission in this spectral band at least 
in obscured sources (Guainazzi \& Bianchi 2007 and references therein). 
Many authors concur to indicate that the origin of the soft X-ray emission in these sources is caused by  photoionisation of extended circumnuclear gas on kpc scale irradiated by the active nucleus which produces He and H-like  transitions of heavy elements and L-shell transitions from Fe (e.g. NGC~1068, Kinkhabwala et al. 2002; Mrk~3, Bianchi et al. 2005, NGC~4151, Armentrout et al. 2007, to cite the most outstanding cases). 

On the other hand, the scenario for unobscured AGN is not so definite.
The lack of line-of-sight obscuring material in this type of sources, allows the whole nuclear 
radiation to escape unblocked to the observer, so that the soft X-ray spectrum is dominated by a smooth excess of continuum rather than  features from reprocessed material (Piconcelli et al. 2005, Crummy et al. 2006). 

In about half of the Seyfert~1 galaxies (Crenshaw et al. 2003, Blustin et al. 2005), the presence of intervening ionised gas is directly observable as a series of absorption features, which present velocity shifts when originating in outflowing or inflowing gas. For the sources studied with the longest integration times, 
it has been possible to place some constraints on the electron density of the gas and, consequently, on  the distance of these warm absorbers, mainly through variability studies.
The emerging picture is far from being a homogeneous description of the circumnuclear ionised gas.
 Among many cases, we recall that the distance of the absorber(s) in  NGC~3783 
was estimated to be   1-3~pc from the nucleus (Netzer et al. 2003 and Behar et al. 2003), whereas it was found to be consistent with subparsec scale in NGC~3516 (Netzer et al. 2002) and it was estimated to be within 4 light days from the nucleus in NGC~4051 (Krongold et al. 2007). 
\begin{table*}      
\centering    
\caption{\label{tab:log} Observation Log of Mrk~335 as observed by {\it XMM-Newton}}.                 
\begin{tabular}{c c c c c c c c}
\\      
\hline\hline                
 OBSID  & Date &  RGS exp &  pn exp  & $\Gamma$$_{soft}$ &  Flux$_{0.3-2}$  & $\Gamma$$_{hard}$ &  Flux$_{2-10}$ \\    
  -           &   (dd/mm/yyyy) & (ks)   & (ks)  &   - &   (10$^{-12}$ ergs cm$^{-2}$ s$^{-1}$)   & - & (10$^{-12}$ ergs cm$^{-2}$ s$^{-1}$)  \\
\hline\hline 
\\ 
0510010701$^{a}$ & 10/07/2007 & 22    & 15  &  2.84$\pm$0.03 & 1.95$\pm$0.04 & 1.02$\pm$0.07   & 3.34$^{+0.23}_{-0.10}$  \\

0306870101$^{b}$ & 03/01/2006 & 130  &  92 & 2.72$\pm$0.01 & 32.16$\pm$0.04 & 2.09$\pm$0.01  & 17.72$\pm$0.07 \\                 

0101040101$^{c}$ & 25/12/2000 & 34    & 30  & 2.80$\pm$0.01 & 31.74$\pm$0.06 & 2.16$\pm$0.02  & 15.00$\pm$0.02 \\
 \hline                                  
\end{tabular} \\
a) Grupe et al. (2008), b) O'Neill et al. (2007), c) Longinotti et al. (2007). The X-ray properties have been measured from the pn spectra. 
\end{table*}

Mrk~335 is a Seyfert 1 galaxy ({\it z}=0.026) with a long X-ray history, since it has been observed by almost all X-ray observatories in the past.
 {\it EXOSAT} (Turner \& Pounds, 1989) and {\it  BBXRT} (Turner et al. 1993a) revealed the presence of an excess of emission in the soft X-rays band, later confirmed in {\it ASCA} data by  Reynolds (1997).
No  clear evidence of warm absorption was established (Reynolds, 1997; George et al. 1998), but {\it ROSAT} data tentatively suggested spectral complexity in the soft X-ray  (Turner et al. 1993b). Nandra \& Pounds (1994) found evidence of a hard X-ray edge  in the {\it Ginga} data.
The presence of the Compton reflection component and a strong Fe K$\alpha$ line was revealed in {\it BeppoSAX} data (Bianchi et al. 2001) and later confirmed in {\it XMM-Newton} observations (Gondoin et al. 2002, Longinotti et al. 2007, O'Neill et al. 2007).
In 2006 the source was observed by the {\it Suzaku} satellite (Larsson et al. 2007).
The spectrum was characterised by a strong soft excess and a reflection component, closely resembling  the {\it XMM-Newton} data; but despite the availability of the high-energy data up to 40~keV,
it has not been possible to model the broadband spectrum unambiguously.   
In May 2007, {\it Swift}  caught Mrk~335 in a historical low X-ray flux state,
with an extremely hard X-ray power law above 2~keV  (Grupe et al. 2007). 
 Following this discovery, the source was observed again by {\it XMM-Newton} in July 2007 as a target of opportunity (ToO) (Grupe et al. 2008).
 This paper reports on the high-resolution spectrum obtained by the Reflection Grating Spectrometer (RGS) onboard  {\it XMM-Newton}.

\section{Observation and data reduction}
Mrk~335  was observed  by {\it XMM-Newton} on 10  July 2007 (OBSID 0510010701) for a duration time of 25~ks. 
Data from the three instruments EPIC (0.3-10~keV), RGS (5-38~$\AA$), and Optical Monitor (OM, 6 filters in the 1800-6000~$\AA$ bandpass)  were available (Struder et al. 2001; Den Herder et al.  2001; Mason et al. 2001).
The detailed analysis of EPIC and OM data will be published in Grupe et al. (2008).
In this paper, only  the RGS data from the ToO observation are considered.
However, Table~\ref{tab:log} reports the broadband properties from all the three {\it XMM-Newton} available data sets to allow the reader an immediate comparison among the flux states of this source.
The raw data were processed with SAS 7.1.0 with the task \texttt{rgsproc} for the RGS data,
which produces spectral products for the source and the background and response matrices.
No background flares due to high-energy particles are present in this observation.
For the OM data, we decided to extract the magnitudes and fluxes from the pipeline products (PPS).

\section{RGS spectral analysis}
\begin{figure*}
  \centering
   \includegraphics[height=1.1\textwidth,width=0.5\textwidth,angle=-90]{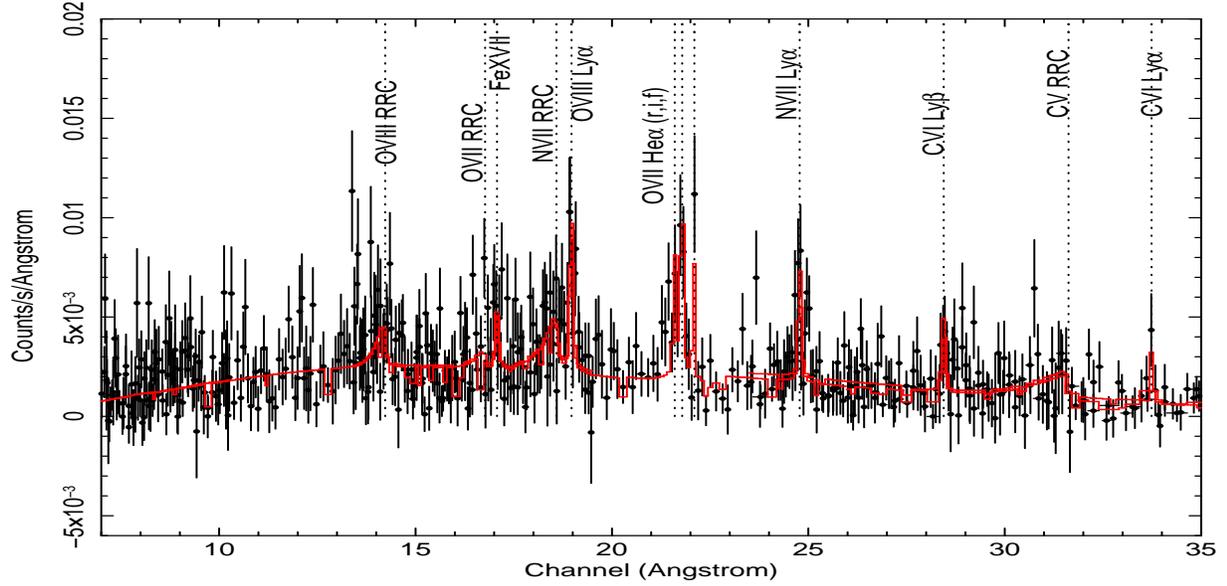}
   \caption{The RGS spectrum of Mrk 335 (corrected for the source redshift).
   The fitted model, plotted as a red line, consists of a power law with $\Gamma$$\sim$2.7 
and the emission features reported in Table~\ref{table:lines}. The data have been  binned according to the instrumental resolution only for plotting purposes (see text).}
              \label{fig:rgs_spectrum}
    \end{figure*}
The spectral analysis was performed using the fitting package SPEX\footnote {http://www.sron.nl/divisions/hea/spex/version2.0/release/index.html}(ver. 2.0),  simultaneously fitting both spectra from the two RGS cameras.  A preliminary analysis was carried out using the first-order spectra, re-binned according to the resolution of the instrument and with a signal-to-noise ratio of 5 over the whole 5-38~$\AA$ range.
This binning was used only to test the agreement between  the power-law slopes of the RGS and of the EPIC data  and to visually check for the presence of broad features in the spectrum (Fig.~\ref{fig:rgs_spectrum}). 
Otherwise, since many spectral channels fall in the limit of low number of photons,
 the unbinned spectrum was used for the spectral analysis  and the C statistic was applied (Cash, 1979).
The quoted errors correspond to the 1$\sigma$ level (i.e. $\Delta$C=1, for one interesting parameter).
Galactic absorption of column density N$_H$=3.9$\times$ 10$^{20}$cm$^{-2}$ is included in all the models (Dickey \& Lockman, 1990). 
\begin{table} 
\caption{\label{table:lines} Fluxes of the soft X-ray lines found in the RGS spectrum.}
\begin{tabular}{c c c c c}
\hline\hline                
Transition & $\lambda$ (lab) & Flux$^1$$_{FWHM=0}$  & $\Delta$C  & Flux$^2$$_{FWHM(OVIII)}$ \\     
-          &  ($\AA$)        &  ph m$^{-2}$ s$^{-1}$  &     -         & ph m$^{-2}$ s$^{-1}$ \\
\hline\hline
\\
 FeXVII             &    17.073  &  $<$~0.05        & $\Delta$C=1  &   0.06$\pm$0.04    \\   
OVII He$\beta$      &    18.627  &  $<$~0.03        & $\Delta$C=0  &  0.06$\pm$0.04    \\         
OVIII Ly$\alpha$    &    18.969  &  0.20$\pm$0.04   & $\Delta$C=31 &  0.27$\pm$0.05   \\  
OVII He$\alpha$ (r) &    21.600  &  0.10$\pm$0.06   & $\Delta$C=4  &  0.17$\pm$0.08    \\
OVII He$\alpha$ (i) &    21.790  &  0.26$\pm$0.08   & $\Delta$C=21 &  0.31$\pm$0.09    \\
OVII He$\alpha$ (f) &    22.101  &  0.17$\pm$0.07   & $\Delta$C=12 &  0.16$\pm$0.07     \\ 
NVII    Ly$\alpha$  &    24.781  &  0.13$\pm$0.05   & $\Delta$C=12 &  0.18$\pm$0.05     \\
CVI Ly$\beta$       &    28.446  &  0.13$\pm$0.06   & $\Delta$C=13 &  0.20$\pm$0.07     \\
CVI Ly$\alpha$      &    33.736  &  0.15$\pm$0.06   & $\Delta$C=8  &  0.24$\pm$0.07       \\
 \hline\hline                                   
\end{tabular}
 1) Line width fixed to FWHM=0;  2) Line width fixed to FWHM=2200~km~s$^{-1}$.   
The energy centroids were fixed to the laboratory value in both cases. The column with the improvement in terms of $\Delta$C refers to the detection significance for the fit with the zero-width lines. 
\end{table}

At  first glance, the RGS spectrum of Mrk~335 appears to be characterised by a weak continuum and a number of emission features (Fig.~\ref{fig:rgs_spectrum}).
The continuum has been fitted with a power law with $\Gamma$=2.75$\pm$0.17, 
in agreement with the slope measured by the pn instrument (Table~\ref{tab:log}).
 All the transitions listed in Table~\ref{table:lines} have been included in the model to fit the narrow emission lines in the spectrum.
Each line was fitted with a delta line model, fixing the centroid energy to the laboratory values. Lines fluxes  are reported in Table~\ref{table:lines}. All but two lines, for which only upper limits are found, yield an improvement in the fit higher or equal to $\Delta$C=4, corresponding to 95.4\% for one free parameter. The final fit statistic including all transitions in Table~\ref{table:lines} is C-statistic= 6247 for 5264 degrees of freedom (d.o.f.).
As a second step, we checked for the presence of the radiative recombination continua (RRC) from CV, NVII, OVII, and OVIII at $\lambda$=31.622, 18.587, 16.771, and 14.228~$\AA$, which are visible in the binned spectrum.
They were incorporated in the model, but when this is fitted, 
only upper limits on the emission measures could be found, so they are not considered in the following. A careful visual inspection of the spectrum did not reveal any absorption feature.

The detected lines were then  checked for line broadening.
 The  OVIII~Ly$\alpha$ line is the most intense isolated transition in the spectrum, 
 and it is not affected by bad pixels.
A Gaussian profile was fitted to this line with the wavelength frozen to the expected value of  18.96~$\AA$ and the width free to vary.
The measured line FWHM is 0.14$\pm$0.05~$\AA$.
We fitted all the spectral lines fixing their FWHM  to the value corresponding to the one of the OVIII~Ly$\alpha$, obtaining the fluxes 
reported in the third column of Table~\ref{table:lines}.
The spectrum was also checked for line shifts by leaving the centroids of the strongest transitions free to vary.
For the OVIII~Ly$\alpha$ line, a blueshift of $\Delta$$\lambda$=-0.014$\pm$0.013~$\AA$ is found, with an improvement of $\Delta$C=7. For the OVII intercombination line the 90\% limits on the shift with respect to the laboratory wavelength are +0.018 and -0.054~$\AA$.
Taking into account that the present signal-to-noise ratio does not allow us to distinguish line contamination and that the RGS systematic error of 8~m$\AA$ is included in the error estimate, we conclude that the peak wavelengths are consistent with the laboratory  values.

\subsection{A close look at the H-like and He-like emission lines in the spectrum}
\begin{figure}
   \includegraphics[angle=90,width=8cm]{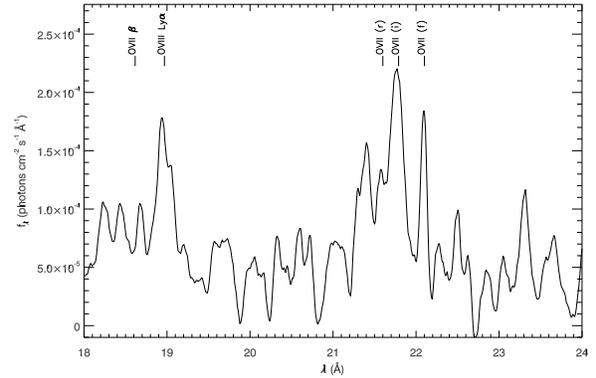}
   \caption{Zoom of the RGS spectrum in the energy range  containing the 
   emission lines from oxygen.}
              \label{fig:triplet}
    \end{figure}

The spectral fit in the previous section and the fluxes reported in Table~\ref{table:lines}
reveal that the  OVII triplet is characterised by dominant intercombination line emission and 
 weaker resonant and forbidden lines (see Fig.~\ref{fig:triplet} for a zoom on this spectral portion).
The line ratios in He-like ions  provide  a powerful diagnostic of the physical properties 
of the emitting gas, mainly by indicating whether the X-ray lines are produced  by 
collisional plasma, by photoionisation, or in a hybrid environment  (Porquet \& Dubau 2000; Porter \& Ferland  2007). 
The lines ratios from the present spectra were calculated with the  OVIII~Ly$\alpha$ and the OVII triplet  lines.
  The numbers estimated  are $\displaystyle\frac{OVIII~Ly\alpha}{OVII(f)}=1.17\pm0.53$,
 G=$\displaystyle\frac{(f+i)}{r}$=4.30$\pm$2.70, L=$\displaystyle\frac{r}{i}$=0.38$\pm$0.25, and R=$\displaystyle\frac{f}{i}$=0.65$\pm$0.33, where {\it r,i,f} indicate the intensity of the resonant, intercombination, and forbidden  line components  in the triplet.
 Given the high value of the G ratio, albeit measured here with considerable uncertainties, the reported line ratios are consistent with an origin in photoionised gas  and a possible contribution from collisional processes (Porquet \& Dubau 2000).

Extensive simulations were run through the photoionisation code 
 {\small CLOUDY} (Ferland et al. 1998)  in order to constrain the physical properties of the emitting gas.
The aim of  these simulations was to extract the expected theoretical values for the line fluxes and to compare them  to the observed fluxes.
 In addition to the line ratios cited above, we also consider the following ratios:
 $\displaystyle\frac{CVILy\alpha}{OVIIILy\alpha}=0.75\pm0.33$,  $\displaystyle\frac{CVILy\beta}{OVIIILy\alpha}=0.65\pm0.33$, 
 $\displaystyle\frac{NVIILy\alpha}{OVIIILy\alpha}=0.65\pm0.28$, and $\displaystyle\frac{CVILy\beta}{CVILy\alpha}=0.86\pm0.53$.
These line ratios were calculated assuming the fluxes from the first column in Table~\ref{table:lines} i.e. 
assuming lines with a delta profile. The fluxes measured with non zero-width Gaussian profiles are nonetheless consistent with this estimate.

  The theoretical  line fluxes were calculated  assuming the spectral energy distribution (SED)
 of a standard AGN as calculated by Korista et al. (1997).
 The treatment of the UV radiation field is particularly important 
 when dealing with a high ratio between the intercombination and the forbidden lines.
In fact, an intense UV source can provide enough photons to depopulate the forbidden level 2~$^3$S via photo-excitation and then pump the electrons into the intercombination level  2~$^3$P, resulting 
in a more intense intercombination line (Mewe \& Schrijver 1978; Kahn et al. 2001).
We  therefore compared the observed SED of Mrk 335 to the one assumed by {\small CLOUDY}. The UV spectrum as observed by the {\it XMM-Newton} Optical Monitor is in good agreement with the SED by Korista et al. (1997).  The spectral slope of the UV-X-ray power-law commonly defined as 
 $\displaystyle\alpha_{ox}=-0.3838~log\left[\frac{F_{\nu}(2keV)}{F_{\nu}(2500\AA)}\right]$ 
was estimated using the pn and OM fluxes at 2~keV and at 2340~$\AA$, respectively giving F$_{2keV}$=0.53$\times$10$^{-11}$~ergs~cm$^{-2}$~s$^{-1}$~keV$^{-1}$ and F$_{2340}$=2.5$\times$10$^{-14}$~ergs~cm$^{-2}$~s$^{-1}$~$\AA$$^{-1}$. 
 The resulting  UV-X-ray power law $\alpha$$_{ox}$=-1.32 is fully consistent with the average Seyfert value assumed by {\small CLOUDY}, i.e. $\alpha$$_{ox}$$\sim$1.4.
  \begin{figure}
 \centering
 \includegraphics[width=9.cm,height=7cm]{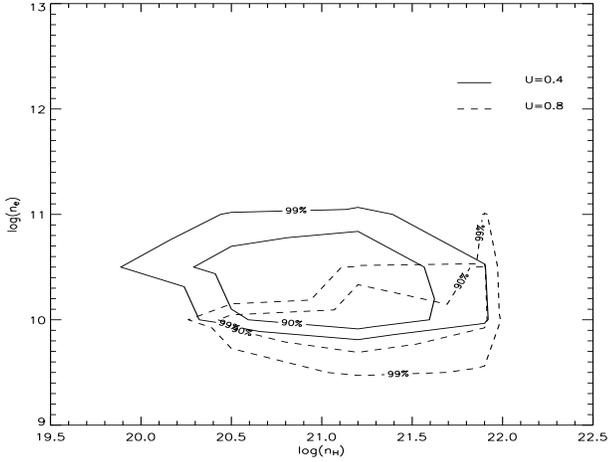}
 \caption{Contour plots for electron and column density of the line emitting gas. The contours  were obtained by comparing the ratios of the emission lines listed in Table~\ref{table:lines} to the values predicted by {\small CLOUDY} simulations (see text for details). The curves define the 90 and 99\% regions  in the parameters space for {\it for all the eight line ratios}. Only  those values of the ionisation parameter yielding a valid solution have been plotted. The corresponding physical quantities are listed in Table~\ref{cloudy}.}
\label{fig:contours}
 \end{figure}

The simulations were carried out by varying the column density of the gas N$_H$ and  ionisation parameter U \footnote{U is defined as $\displaystyle{\frac{\phi(H)}{n_e~c}}$ where the radiation field is expressed by 
$\displaystyle{\phi(H)=\frac{k}{4\pi~r^2_0}\int^{\nu2}_{\nu1}}$~$\displaystyle{\frac{\pi~F_{\nu}}{h\nu}d\nu}$ with $\nu1$=1Ryd and $\nu2$=$\infty$ and the relation between the normalisation costant {\it k}, and the bolometric luminosity of the source is expressed by $\displaystyle{L=k\int^{\nu2}_{\nu1}\pi F_{\nu}d\nu}$.}
over a range of electronic density n$_e$ spanning  10$^6$ to 10$^{14}$~cm$^{-3}$.
These quantities were varied in small steps, so that a fine grid of possible solutions was obtained.
In particular, the ionisation parameter is stepped between logU=-2 and log U=2.

 To compare the measured line ratios to the spectral model simulated by {\small CLOUDY}, each line ratio  was searched  over the grid of predicted  values in the following way.
 For each line ratio {\it i} the observed value L$_{o,i}$ with its error err(L$_{o,i}$) and the corresponding 
theoretical value L$_{t,i}$ have been taken into account and used to calculate the quantity
$\sigma^2_i$=$\displaystyle{\frac{(L_{o,i}-L_{t,i})^2}{err(L_{o,i})^2}}$

Then, the quantity $(\sum\sigma^2_i)^{1/2}$  has been minimised,  {\it i} being the summation index over the considered line ratio.
The contour plots in the N$_h$-n$_e$ plane have been drawn by imposing a distance equal to 4.61 and 9.21 from the minimum solution, corresponding to 90 and 99\%  levels of confidence. The inner and outer contours in Fig.~\ref{fig:contours} therefore represent the locii of valid solutions distributed respectively at 90 and 99\%  considering the eight line ratios simultaneously.

The measured data points are consistent only with {\small CLOUDY} solutions
 in the  range  logU=0.4--0.8, corresponding to the continuous  and dashed lines in Fig.~\ref{fig:contours}.
Since one of the output of  {\small CLOUDY} is the column density of the gas,
   an estimate of the size of the emitting material  {\it l} was provided using the relation  {\it l}=N$_H$/n$_e$.
For each solution, the distance {\it r$_0$} of the cloud from the ionising source  can be extracted from {\small CLOUDY} by replacing the AGN bolometric luminosity in the definiton of the  ionisation parameter given above.  
The value  L$_{bol}$=10$^{44.7}$~ergs~s$^{-1}$ was adopted after Woo \& Urry (2002).
For each ionisation parameter U, the  values of the distance of the emitting gas cloud(s) from the nucleus and its size have been estimated. 
 The minimum and maximum values correspond to the 99\% contour for the range in electron and column density of the gas in Fig.~\ref{fig:contours}.
The solutions admitted by our data are summarised in Table~\ref{cloudy} and they will be discussed in the following section.
\begin{table}       
\centering 
\caption{\label{cloudy}  Summary of the physical properties of the gas clouds estimated from  the 99\% contour plot in Fig.~\ref{fig:contours}.}  
\begin{tabular}{c c c c c }
\hline\hline                
 logU & {\it n$_e$ range}    & {\it  n$_H$ range}    &  {\it dist range} &  {\it size range}  \\     
        - & cm$^{-3}$      &   cm$^{-2}$         &     cm         &         cm      \\
\hline\hline
\\
 0.4  &  10$^{9.8-11.1}$  &  10$^{19.9-21.9}$     &   1.95$\times$10$^{16-17}$  &   10$^{8-12}$    \\
 0.8  &  10$^{9.5-11}$  &  10$^{20.3-22}$     &   1.95$\times$10$^{16-17}$  &   10$^{9-13}$      \\
  \hline\hline                                   
\end{tabular}
\end{table}

%

\section{Discussion}
The RGS spectrum of Mrk~335 at low state provides  a unique opportunity for getting insights into an X-ray reprocessing region that  is not easily accessible in Seyfert~1 objects.  
The soft X-ray emission lines could be observed because of the decrease in the continuum flux.  The nuclear power was attenuated for reasons as yet unknown, however the most straightforward explanation could be partial obscuration of the central source.
At first glance, this object seems to be analogous to the other well-known case of an extremely variable 
Seyfert~1, NGC~4051, which showed a very rich emission line spectrum  when it  was observed in a  low flux state by {\it XMM-Newton} (Pounds et al. 2004).
According to these authors, the analysis of the emission lines in this source, especially of the OVII triplet, pointed to interpret the soft X-ray spectral features as arising from photoionised low-density gas distributed on a large scale,  consistent with the AGN narrow line region.
The variability history of  Mrk~335  and the comparison with other variable AGN are not the prime focus of the present paper.
 However, the similarity between NGC~4051 and Mrk~335 
suggests to consider first the geometry of the system.
 \begin{figure}
 \centering
\includegraphics[width=\columnwidth]{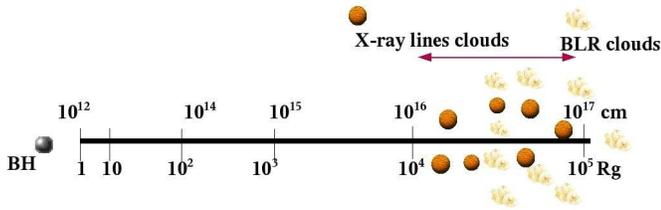}
\caption{The cartoon shows qualitatively the location of the ionised gas in the nucleus of Mrk~335 on a scale of gravitational radii and the corresponding cm scale. }
\label{fig:nucleus}%
 \end{figure}

Mrk~335 is a pure Seyfert~1 galaxy, since the nuclear emission does not suffer {\it any} obscuration nor does absorption when the source is observed at high state, as shown by previous, recent X-ray observations (Longinotti et al. 2007, O'Neill et al. 2007). Nonetheless, the low-state, soft X-ray spectrum resembles the RGS spectra recently observed in obscured AGN by  Guainazzi \& Bianchi (2007).
 It is then reasonable  to postulate that we are observing the same line-emitting gas in Mrk~335 as became observable  due to the drop in the continuum flux. 
 There are some objections to this hypothesis.
   The main point is that, in many obscured sources, the soft X-ray  photoionised matter was demonstrated to be coincident with the extended narrow line region, i.e. well out of the obscuring torus (Bianchi et al., 2006).
The scale  of this gas is much larger than the distance of the emitter observed in Mrk~335
for  all the possible solutions listed in Table~\ref{cloudy}.
For this reason,  the hypothesis of a common location of the soft X-ray 
photoionised gas for unobscured and obscured objects is discarded in the present case.

The simulations run with {\small CLOUDY} cannot uniquely constrain the ionisation 
parameter or the electron density of the gas, and yet they provide a range 
of physical conditions for the gas that are noticeably in agreement. 
The distance of the photoionised plasma from the nuclear source
ranges approximately  from 7 to 77 light days, meaning that the gas is always 
confined within $\sim$0.06 pc.
The column density never gets higher than 10$^{22}$~cm$^{-2}$ (see contours in Fig.~\ref{fig:contours}) and the size of the emitting region(s) always appears quite small, the dimension of the largest cloud being $\sim$10$^{11}$~cm. 
From the black hole mass M=14.2$\pm$3.7$\times$10$^{6}$~M$\odot$ (Peterson et al. 2004), 
the Schwarzschild radius R$_s$ is estimated to be around 4$\times$10$^{12}$~cm.
The distance of the Broad Line Region (BLR) from the central source can be found by considering 
the width of the optical lines and assuming that the BLR gas is virialised.
 The H$\beta$~FWHM was measured by Boroson \& Green (1992), and it is equal to 
 1640~km~s$^{-1}$, in agreement with the reverberation mapping measurements provided by 
  (Peterson et al. 2004).
The BLR can then be located at $\sim$7$\times$10$^{16}$~cm, and that this gas in Mrk~335 is constituted of several clouds of emitting material forming a clumpy medium was suggested by  Arav et al. (1997) on the basis of observational constraints. 
  
It is very likely that the X-ray photoionised gas is inner to the BLR clouds and it is also reasonable  that the two materials form a continuous distribution of clouds.
One can imagine that the innermost clouds spread out towards the nucleus, reaching an 
ionisation level high enough to emit the observed soft X-ray lines.
In this case, the spectral line should suffer line broadening due to the vicinity 
of the central black hole, although this effect should not be extreme because the Schwarzschild radius is 3 orders of magnitude 
smaller.
The Gaussian line width measured in the OVIII Ly$\alpha$  corresponds to a FWHM of 2200$\pm$750~km~s$^{-1}$. We checked for the width in the OVII intercombination line, and the upper limit on the line width is consistent with the measurement from the OVIII~Ly$\alpha$ at $<$0.15~$\AA$.  
 The distance estimated from the virial assumption is then constrained within  $\sim$2.3$\times$10$^{16}$~cm and 1.2$\times$10$^{17}$~cm. 
In principle, this value can be easily reconciled with the distance inferred from {\small CLOUDY}; on the other hand, it may add an independent observational constraint on the structure of the gas. In fact, the underlying assumption in our {\small CLOUDY} model is that all the X-ray emission lines originate in the same cloud, therefore the distances in Table~\ref{cloudy}  must  refer  to the bulk of the photoionised gas.
The distance estimated from the OVIII~Ly$\alpha$ width may indicate that the 
spectral lines come from a distribution of clouds and that OVIII is only concentrated  in the innermost ones. To conclude, the X-ray photoionised gas is likely to be located within  the optical BLR, as depicted qualitatively in Fig.~\ref{fig:nucleus}.
 Longer exposures  of this AGN at a serendipitous low state are needed to reach firm conclusions.

 Finally, it has to be remarked that the existence of circumnuclear ionised gas in the innermost region  is not a novelty in this AGN.
 Longinotti et al. (2007) reported on the detection  of a narrow absorption feature around  5.9~keV in the EPIC spectrum of the 2000 observation (see Table~\ref{tab:log}). This feature was identified as a redshifted Fe~XXVI K$\alpha$ transition and interpreted as the signature of  gas inflowing towards the nucleus at a velocity of 0.11-0.15~{\it c}.
Six years later, {\it XMM-Newton} observed the source again; the detection of an {\it emission} line 
at $\sim$7~keV  in the EPIC spectrum was associated to the presence of 
 highly ionised material on a distance scale much larger than those discussed herein (O'Neill et al. 2007).  
 Interestingly, the most recent X-ray observations of Mrk~335 performed during the past year by {\it Suzaku}
  and {\it XMM-Newton} did not show any emission line at 7~keV (Larsson et al. 2007, Grupe et al. 2008), implying that rapid line variability must be involved.
The potential  physical connection among the briefly outlined observational results and the 
soft X-ray lines  in the present work should be a point of major interest in the next studies of this intriguing active galaxy.

\begin{acknowledgements}
This paper is based on observations obtained with XMM-Newton, an ESA science mission with instruments and contributions directly funded by ESA Member States and NASA.
The authors wish to thank  the XMM-Newton Project Scientist Norbert Schartel for coordinating the observation and the release of the {\it XMM-Newton} data.  
We are grateful to many people within the XMM Science Operation Centre: 
Pedro Rodriguez-Pascual for his contribution on Mrk~335 Optical/UV SED,  
Andy Pollock for always being a wise advisor on the RGS data treatment, and 
 Maria Diaz-Trigo for support with the use of the SPEX software.
 The authors thank R. Porter  for  help with  {\small CLOUDY} simulations.  
  We thank Dirk Grupe for a fruitful exchange of opinions on the XMM-Newton data, and 
we warmly thank Yair Krongold for many stimulating discussions
on this paper during his visit at ESAC.
 \end{acknowledgements}

\end{document}